\documentclass[prc,superscriptaddress,preprintnumbers,unsortedaddress,twocolumn,dvips]{revtex4}

\usepackage{amsmath, amssymb}
\def\la{\mathrel{\mathpalette\fun <}}
\def\ga{\mathrel{\mathpalette\fun >}}
\def\fun#1#2{\lower3.6pt\vbox{\baselineskip0pt\lineskip.9pt
\ialign{$\mathsurround=0pt#1\hfil##\hfil$\crcr#2\crcr\sim\crcr}}}

\usepackage{graphicx}

\usepackage{times}
\usepackage{color}
\usepackage{ulem}
\usepackage{bm}

\begin{document}

\preprint{NITEP 45}

%---- title and author ----
\title{Nuclear medium effect on neutron capture reactions during neutron star mergers}

\author{Kazuyuki Ogata}
\email[]{kazuyuki@rcnp.osaka-u.ac.jp}
\affiliation{Research Center for Nuclear Physics (RCNP), Osaka University, Ibaraki 567-0047, Japan}
\affiliation{Department of Physics, Osaka City University, Osaka 558-8585, Japan}
\affiliation{Nambu Yoichiro Institute of Theoretical and Experimental Physics (NITEP), Osaka City University, Osaka 558-8585, Japan}

\author{Carlos A. Bertulani}
\email[]{carlos.bertulani@tamuc.edu}
\affiliation{Department of Physics, Texas A\&M University, Commerce, TX 75429, USA}

\date{\today}

\begin{abstract}
The discovery of gravitational waves has confirmed old theoretical  predictions that binary systems formed with compact stars play a crucial role not only for cosmology and nuclear astrophysics \cite{LS74}. As a byproduct of these and subsequent observations, it is now clear that neutron-star mergers can be a competitive site for the production of half of the elements heavier than iron in the universe following a sequence of fast neutron capture reactions known as the r process. In this article we discuss an effect which has been so far neglected in calculations of r-process nucleosynthesis in neutron star mergers. We show that the corrections due to the neutron environment even at relatively small neutron densities, within the bounds of numerical hydrodynamical  simulations of neutron star mergers and after the onset of the r process, are non-negligible and need to be taken into account to accurately describe the elemental abundance as determined by observations.
\end{abstract}

\maketitle

{\it Introduction.}
The rapid neutron capture process (r process) is thought to be responsible for about half of all elements heavier than iron. However, the properties of the nuclei and nuclear reactions involved in the r process are not well known. Besides, its astrophysical site is not yet clearly identified. Neutron star mergers and supernova explosions are possible stellar sites candidates for the r process, but it is uncertain which of the two sites are more effective \cite{LS74,Cot17,Nis17}. Neutron star mergers are promising nucleosynthesis sites of heavy elements, because a large amount of neutron-rich matter is probably ejected during merger \cite{LS74,Cot17,Nis17,Eich89,Thie17,Dav96,Ruf97}. For neutron star mergers, the ejecta medium has an electron fraction, or electron number density per baryon density, in the range $0.05 \lesssim Y_e \lesssim 0.5$, irrespective of the equation of state (EoS). This might explain the r-process abundance pattern for nuclear mass numbers $A\gtrsim 90$ observed in the solar system and in metal-poor stars. In the case of black-hole (BH) and NS mergers, the dynamical ejecta contains a low electron number density, $Y_e \lesssim 0.1$, and  heavy elements $A\gtrsim 130$ are produced in the r process \cite{Full17}. But the r process can  create all elements up to $A\sim 250$, beyond the third peak of elemental abundance. It is also possible that very heavy nuclides with $A>300$ can be produced due to the very neutron-rich environment. Those nuclides are unstable to spontaneous or neutron-induced fission leading to a fission cycle  that continues until the free neutrons are exhausted \cite{LS74,Cot17,Nis17,Eich89,Thie17,Dav96,Ruf97,Full17}.

In order to calculate the nucleosynthesis in NS-NS or BH-NS ejecta scenarios one needs to know basic nuclear properties, such as nuclear masses, $\beta$-decay lifetimes, fission properties, (n,$\gamma$) and ($\gamma$,n) reaction rates, and a solid knowledge of nucleon capture cross sections is needed. A novel phenomenon considered in this letter is the influence of the neutron density on the (n,$\gamma$) and ($\gamma,$n)  rates. We show that a substantial reduction of theses cross sections emerges when the r-process nuclei are immersed in a neutron background even for the moderately low densities in the ejecta of the NS mergers.  In fact, simulations have shown that neutron densities up to about $\rho \sim 10^{-3}$--$10^{-2} \rho_0$ are possible during the nucleosynthesis process following the merger stage, where $\rho_0=0.16$ fm$^{-3}$ is the nuclear saturation density \cite{Almud,Frei99,Baus13,Rad17}. The consequences of  the neutron background on the reaction rates have not been studied so far. But a full consideration of this phenomenon, using a hydrodynamical simulation coupled to a nucleosynthesis reaction network path, is beyond the scope of the present work. Our objective is to explore how the presence of the surrounding neutron medium will impact the cross sections and, if the impact is shown to be important, it certainly needs to be considered in a full r-process calculation of NS-NS or BH-NS mergers.

It should be mentioned, however, that the neutron capture rates at the very high densities considered here are probably irrelevant for r-process network calculations. The decompressing neutron star material ejected from the merger is initially quite dense, but its temperature is sufficiently high that nuclear statistical equilibrium is probably achieved. As the temperature and density of the ejected material drop to too low values, the abundance of reactants will be too small for certain reactions to go fast enough to maintain statistical equilibrium. When the temperature drops below approximately 0.1~GK the statistical equilibrium will fail, at which time the astrophysical simulation with assumption of a uniform neutron distribution, predicts a density of about $10^{-5}$ times the nuclear saturation density. Therefore, our results will only prove to be useful if merger simulations predict non statistical conditions while high densities prevail. This cannot be completely ruled out presently. Notwithstanding, our primary aim is to clarify the role of background neutron in neutron-capture processes, which we believe is potentially important for a better understanding of the entire scenario of the nucleosynthesis in NS-NS or BH-NS mergers.

{\it Theoretical framework.} Our approach is based on a standard nuclear reaction theory. We will consider the total neutron-nucleus induced cross section usually dominated by $\gamma$ emission for low neutron energies. We use the following expression for the total neutron capture cross section in free space:
\begin{eqnarray}
\sigma(E)=
-{K\over E} \left< \chi^{(+)} \Big| W \Big| \chi^{(+)} \right>,
 \label{capt}
\end{eqnarray}
where $K$ and $E$ is the center-of-mass scattering momentum and energy, respectively, and $W$ is the imaginary part of the neutron-nucleus one-body potential $U$.
The neutron scattering (distorted) wave $\chi^{(+)}$ is a solution of
the Schr\"odinger equation for the potential $U$ under the outgoing boundary condition. With the total reaction cross section, one can easily obtain the reaction rate  by a folding of its energy (velocity) dependence with the relative velocities of the neutron and the nuclei in the medium, assumed to follow a Maxwell-Boltzman distribution.

We assume that neutrons from the environment are homogeneously distributed
around the reacting (neutron+nucleus) system. These neutrons will be denoted here by
background neutrons (BN). We further assume that the BN are
not substantially affected by the nucleus. This allows one to estimate
the additional potential, besides the neutron-nucleus potential $U$, that the incoming neutron feels under the influence of the BN,
which we call an {\it environment mean potential} (EMP), as
\begin{equation}
V_{\mathrm{EMP}}\left(  \boldsymbol{r}\right)  =
\int_\Omega A {\cal P}(\boldsymbol{r}-\boldsymbol{r}')
v_{nn}
\left(\left\vert\boldsymbol{r}-\boldsymbol{r}^{\prime}\right\vert \right)
d\boldsymbol{r}^{\prime}.
\label{vemp}
\end{equation}
Here, $v_{nn}$ is a neutron-neutron ($nn$) interaction,
$\Omega$ is a macroscopic volume surrounding the reaction system,
$A$ is the number of neutrons inside $\Omega$, and
${\cal P}(\boldsymbol{r}-\boldsymbol{r}')$ is a probability density that
one finds neutron at $\boldsymbol{r}'$ when
the incoming neutron is loclated at $\boldsymbol{r}$. In terms of these definitions,
${\cal P}\Omega$ is a probability measure of the influence of the BN by the neutrons
within a distance $\boldsymbol{r}-\boldsymbol{r}'$ inside the volume $\Omega$. Due to the
saturation of the nucleon-nucleon force, the calculation of $V_{\mathrm{EMP}}$ converges
quickly for distances $s=|\boldsymbol{r}-\boldsymbol{r}'|$ larger than $\sim 3$ fm.

To evaluate ${\cal P}$ we consider the probability density $P$ to find particle 1 at $\boldsymbol{r}_1$ and particle 2 at $\boldsymbol{r}_2$ when the $A$ lowest (continuum) single particle (s.p.) states are occupied. Using a Slater determinant for these states, one gets
\begin{eqnarray}
P(\boldsymbol{r}_1,\boldsymbol{r}_2)&=&\int \psi^*(1,\cdots A)\psi(1,\cdots A)
d\boldsymbol{r}_3 d\boldsymbol{r}_4 \cdots d\boldsymbol{r}_A \nonumber \\
&=&{1\over A(A-1)} {1\over 2} \sum_{\alpha,\beta} \left| \phi_\alpha (1)  \phi_\beta (2)  - \phi_\beta (1)  \phi_\alpha (2) \right|^2, \nonumber \\
\end{eqnarray}
where $\psi$ is an antisymmetrized $A$-body wave function and $\phi$ denotes
a s.p. state.  $P$ is normalized as $ \int_\Omega P d\boldsymbol{r}_1 d\boldsymbol{r}_2=1$.  The relation between
the two probability densities is
\begin{equation}
{\cal P}({\bf s})=\int_\Omega d{\bf r}_1 P(\boldsymbol{r}_1,\boldsymbol{r}_2).
\end{equation}
We assume that each neutron-neutron pair is in a s-wave scattering state with $S=0$ and $T=1$. At low temperatures, we can also safely assume that all neutron states in the continuum are occupied up to the Fermi momentum $k_{\rm F}$, and  the following relation with the number density $\rho=A/\Omega$ of the BN arises:  $ {4\pi}k_{\rm F}^3/3 ={8\pi^3 A / \Omega}$. Using plane waves for the relative motion between the neutrons,
$\phi_\alpha = (1/\sqrt{\Omega}) \exp{(i\boldsymbol{k}\cdot\boldsymbol{r})}$,
it is straightforward to show that for $A\gg 1$, the probability entering Eq. \eqref{vemp} is given by~\cite{BM69}
\begin{equation}
{\cal P}(s)=
{1\over \Omega}
\left(
1 - \left[ {3\over (k_{\rm F}s)^2}\left( {\sin (k_{\rm F}s) \over k_{\rm F}s} -\cos (k_{\rm F}s)\right)\right]^2
\right),
\end{equation}
which does not depend on where one of the two particles is located but only on their separation distance
$s=\left\vert\boldsymbol{r}-\boldsymbol{r}^{\prime}\right\vert$.

\begin{figure}[t]
\begin{center}
\includegraphics[width=0.48\textwidth]{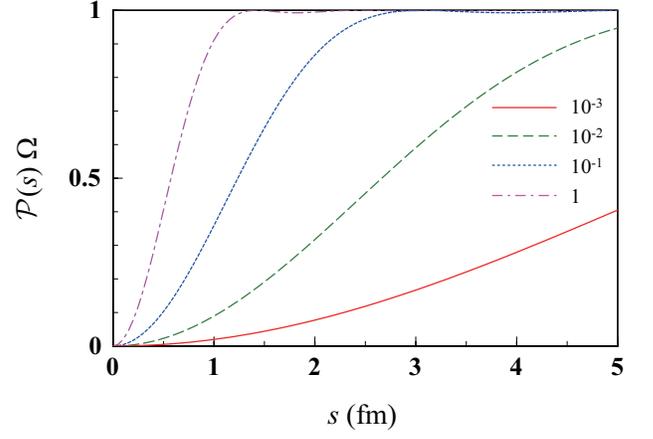}
\caption{The probability due to the environment neutrons, ${\cal P}(s)\Omega$, as defined in Eq. \eqref{vemp},  is shown for four different values of
$\rho/\rho_0=10^{-3}$, $10^{-2}$, $10^{-1}$, and 1 by the solid, dashed-dotted, dashed, and dotted lines, respectively. }
\label{fig1}
\end{center}
\end{figure}

The Pauli blocking probability due to the BN, ${\cal P}(s)\Omega$, is shown in Fig.~1 for four different values of
$\rho/\rho_0=10^{-3}$, $10^{-2}$, $10^{-1}$, and 1 by the solid, dash-dotted, dashed, and dotted lines, respectively.
The Pauli blocking creates an effective {\lq\lq}wound'' at the neutron-neutron scattering wave function for neutron-neutron separation distances  $s < 2/k_{\rm F}$ ($\sim 1.4$ fm at saturation density $\rho_0$). As seen in Fig.~\ref{fig1}, at lower density, the spatial region affected by Pauli blocking becomes wider because of the uncertainty principle. Note, however, that the effect of the BN in the low density limit is negligible because Eq.~(\ref{vemp}) is proportional to $\rho$. If the Pauli principle is ignored, ${\cal P}(s)\Omega =1$ and, equivalently, $A{\cal P}(s) = \rho$. One may notice a slight oscillation of the dash-dotted line. This is an outcome of the Fermi-gas model in which a sharp cutoff of the momentum of each nucleon is made at $k_{\rm F}$.

Integrating  Eq.~(\ref{vemp}), and because of the finite range of the neutron-neutron interaction, one finds that  the net effect of the BN on the neutron-nucleus scattering wave is to effectively shift the scattering energy $E$ by a nearly constant amount,
\begin{equation}
E \rightarrow E+\varepsilon_{\rm EMP}.
\end{equation}
Notice that even in the absence of Pauli blocking, medium effects would have to be present  in the classical limit (${\cal P}(s)\Omega=1$). In this case,
\begin{equation}
\varepsilon_{\rm EMP} \to -\rho \bar{v}_{nn},
\end{equation}
where $\bar{v}_{nn}$ is the volume integral of $v_{nn}$. The effect of Pauli blocking is to reduce this energy shift, allowing the
neutrons to stream more freely within a denser region. This effect is well known in nuclear physics, justifying the
use of the Fermi gas  model to describe some of the gross nuclear properties. The novel idea in our work is
to extend these known features to the case of neutron-nucleus scattering in the presence of a neutron background environment.

According to the energy shift approximation introduced above, the neutron capture reaction rate in a dense neutron environment is corrected to
\begin{eqnarray}
\left\langle \sigma v\right\rangle_{EMP}  &=&
\left(  \frac{8}{\mu\pi}\right)^{1/2}\frac{N_{\rm A}}{ \left( k_{\rm B}T\right)
^{3/2}}\int e^{-E/{k_{\rm B}T}} \sigma\left(  E + \varepsilon_{\rm EMP}\right)
\nonumber \\
&& \times \sqrt{E(E+\varepsilon_{\rm EMP})}dE.
\label{rate2}
\end{eqnarray}
This method allows one to quickly estimate the effects of the BN in a simple fashion, also highlighting its
dependence on the details of the nucleon-nucleon interaction at short distances. A very interesting outcome
of this result is that, at very low energies, when $\sigma \propto 1/v$, where $v$ is the neutron velocity, one gets
$\left\langle \sigma v\right\rangle_{\rm EMP} = \left\langle \sigma v\right\rangle_{\rm free}$. The $1/v$ law emerges directly from Eq. \eqref{capt} since the matrix element on its right asymptotically reaches a constant value as $E\rightarrow 0$. Therefore, at very low energies,  Pauli blocking by the environment neutrons effectively keeps unchanged the reaction rates, as if the  incoming neutron-nucleus reaction energy remains unaltered.

\begin{figure}[t]
\begin{center}
\includegraphics[width=0.48\textwidth]{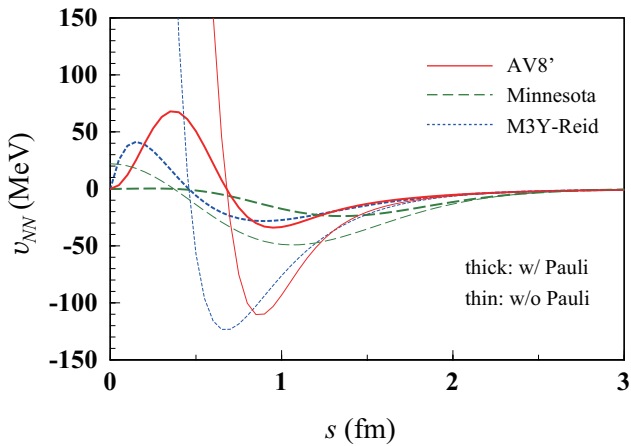}
\caption{The nucleon-nucleon interaction, $v_{nn}$, as a function of the neutron-neutron distance $s$ is shown by the thin lines. The effective interaction $v_{nn}^{\rm eff}$, with the effects of the neutron environment, for $\rho=10^{-1}\rho_0$ is shown by the thick lines. The dotted, solid, and dashed-dotted lines correspond to the AV8', Minnesota, and M3Y-Reid interactions, respectively.}
\label{vnn}
\end{center}
\end{figure}

{\it Application to neutron capture on Ni isotopes.} In the following discussion, we consider neutron capture processes
by $^{58}$Ni and $^{78}$Ni by employing the Koning and Deraloche global optical potential  for $U$, needed to calculate the scattering waves entering Eq. \eqref{capt}.  The choices for these two systems are not special, but just reflecting standard study cases. Also, we would like to check if there is any effect due to an increase of the neutron number of the nucleus. We will conclude that it does not have a substantial change and that the effect of the BN on the reaction rate has very general features common to most neutron-capture reactions in dense environments. Except for the extremely low energies of a few eV (thermal neutrons), there is also a substantial deviation from the $1/v$ law, especially for the energies involved in a stellar environment.  For the neutron-neutron interaction $v_{nn}$ in the $(S,T)=(0,1)$ channel, we use the Argonne  V8' (AV8') interaction, the Minnesota interaction, and the Michigan three-range Yukawa interaction based on the Reid potential (the M3Y-Reid interaction).

In Fig.~2  the nucleon-nucleon interaction, $v_{nn}$, as a function of the neutron-neutron distance $r$ is shown by thin lines. The effective interaction $v_{nn}^{\rm eff}$, with the effects of the neutron environment, for  $\rho=10^{-1}\rho_0$ is shown by thick lines. The solid, dashed, and dotted lines correspond to the AV8', Minnesota, and M3Y-Reid interactions, respectively. One sees that the thin lines largely deviate from each other, while the difference between the thick lines is quite smaller. This is also the case with different values of  $\rho$ down to $10^{-3}\rho_0$. As expected from Fig.~1, the Pauli principle evidently hinders $v_{nn}$ at small distances $s$, where the differences between the three chosen $v_{nn}$ interactions are somewhat large. This is why the interaction dependence is weakened in $v_{nn}^{\rm eff}$. We can therefore expect that the EMP effect introduced in this work does not depend strongly on $v_{nn}$.

\begin{figure}[t]
\begin{center}
\includegraphics[width=0.48\textwidth]{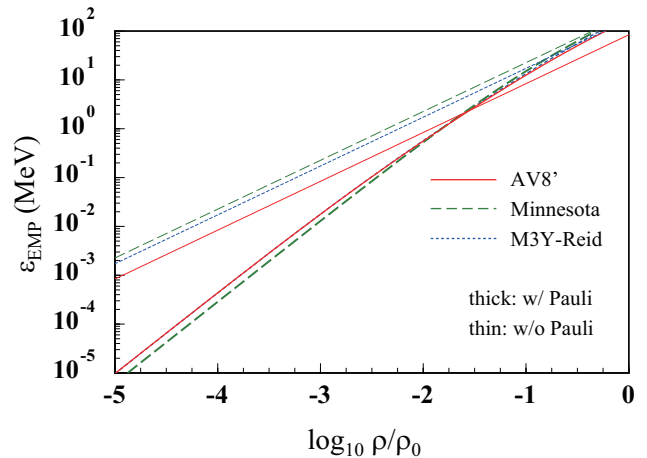}
\caption{Energy shift of the neutron-nucleus scattering energy, $\varepsilon_{\rm EMP}$, as a function of $\log_{10} \rho/\rho_0$. The solid, dashed, and dotted lines correspond to the AV8', Minnesota, and M3Y-Reid interactions, respectively. The thick (thin) lines show the energy shifts when the Pauli blocking is (is not) included.}
\label{DEfig}
\end{center}
\end{figure}

In order to assess the influence of the environment BN we now turn to the neutron-nucleus scattering properties. Figure~3 shows the energy shift of the neutron-nucleus scattering energy, $\varepsilon_{\rm EMP}$, as a function of $\log_{10} \rho/\rho_0$. The meaning of the lines is the same as in Fig.~2.
Comparing the three interactions, the magnitude of the $|\bar{v}_{nn}|$ for the AV8' interaction is found to be smallest because it has the largest repulsive core. Once the Pauli blocking effect is considered (thick lines), this characteristic feature of the AV8' interaction vanishes and the three $v_{nn}$ yield almost the same $\varepsilon_{\rm EMP}$. For the other two $v_{nn}$, Pauli blocking has no significant effect on $\varepsilon_{\rm EMP}$ for $\rho \ga 10^{-1} \rho_0$. At lower $\rho$, the {\lq\lq}wound'' on $v_{nn}$ develops further to larger $r$ and $\varepsilon_{\rm EMP}$ decreases significantly. This is the case also with the AV8' for $\rho \la 2 \times 10^{-2} \rho_0$.
Inclusion of the Pauli principle is thus essential to estimate $\varepsilon_{\rm EMP}$. An important conclusion stemming from Fig.~3 is that there exist a non-negligible energy shift $\varepsilon_{\rm EMP}$ of about 20 (200)~keV even at $\rho/\rho_0= 10^{-3}$ ($5\times10^{-3}$).

\begin{figure}[htbp]
\begin{center}
\includegraphics[width=0.48\textwidth]{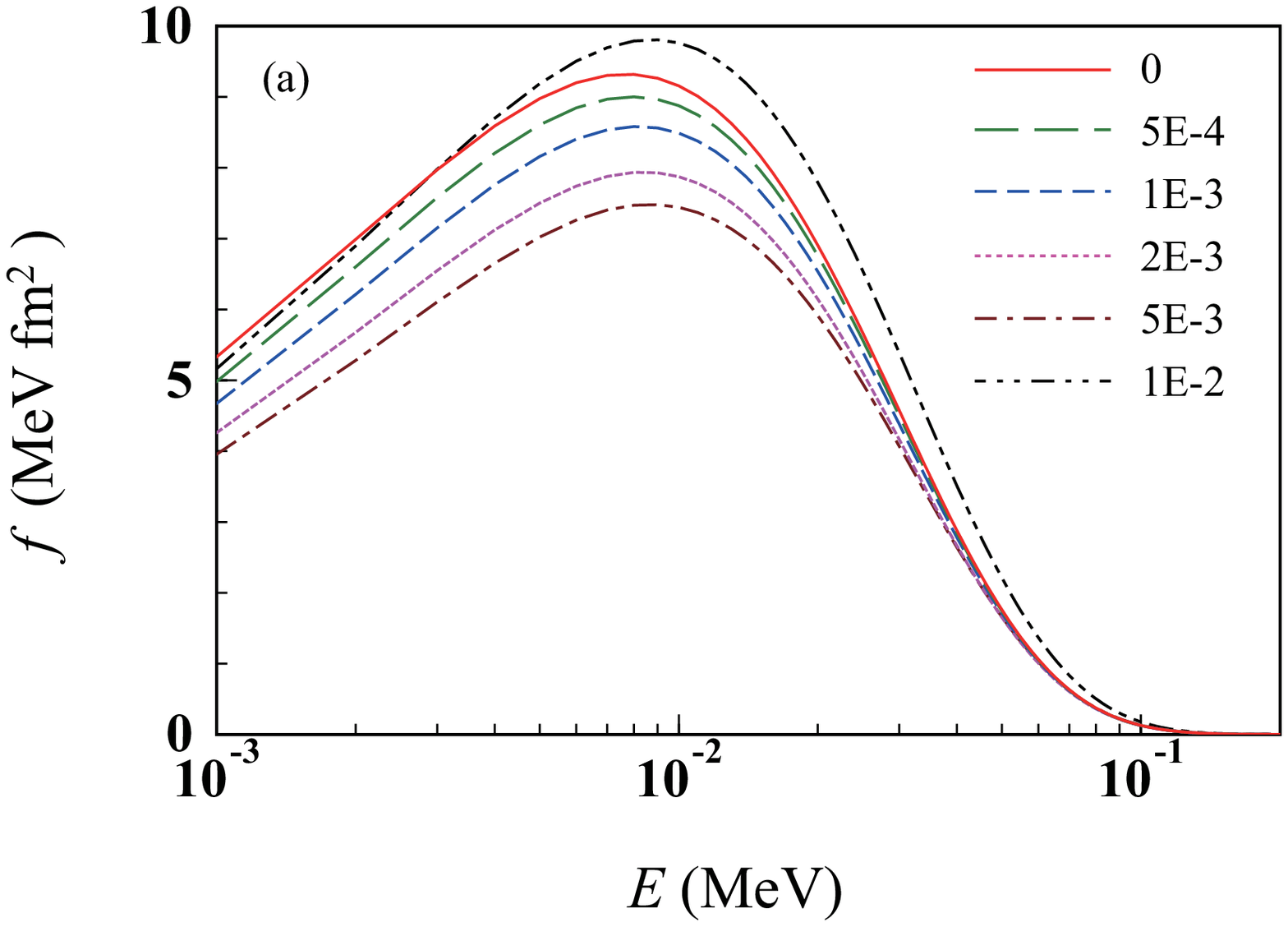}
\includegraphics[width=0.48\textwidth]{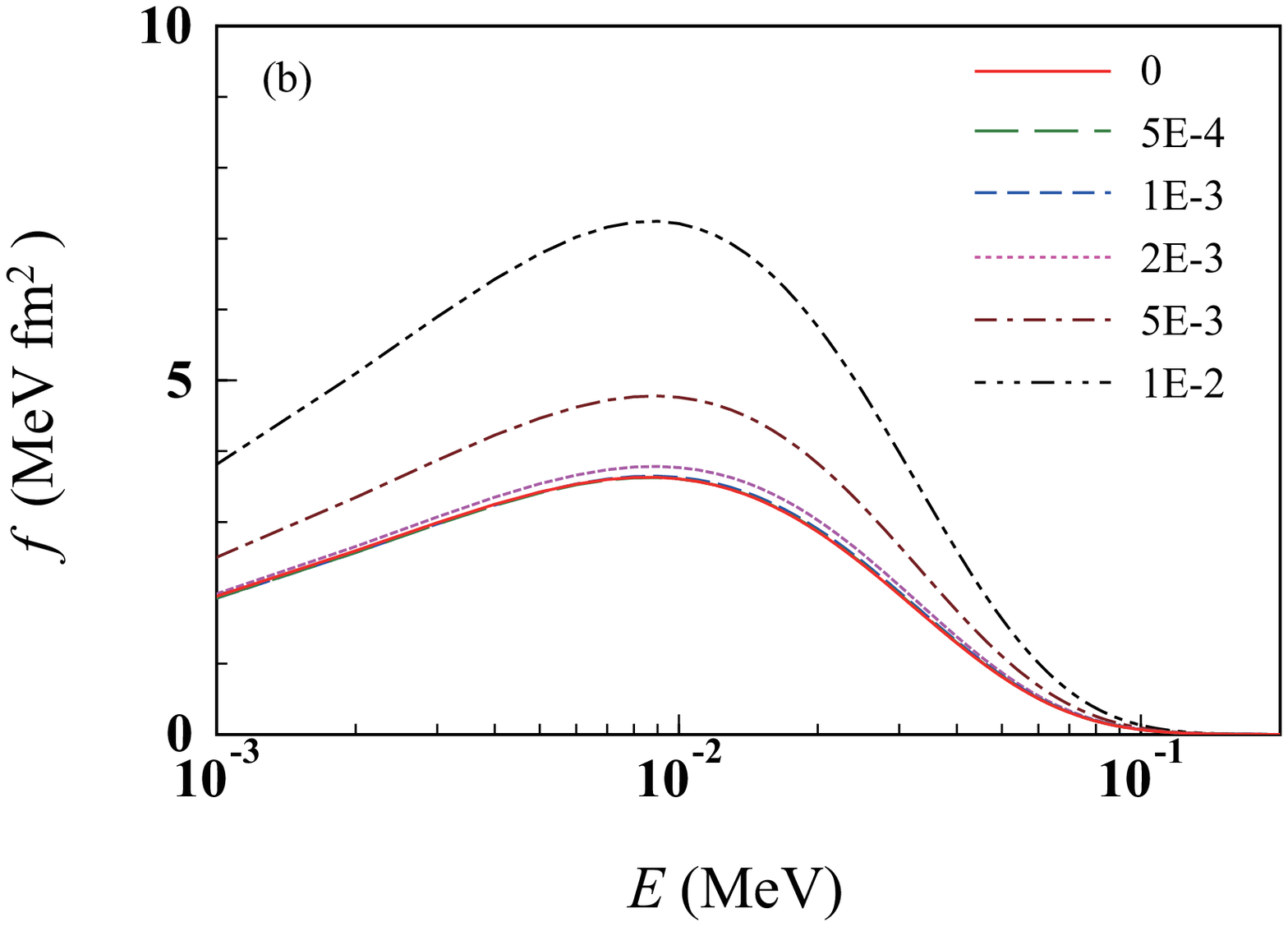}
\caption{The integrand $f$ in Eq.~(\ref{rate2})  (a) for $^{58}$Ni and (b) for $^{78}$Ni,  as a function of the neutron scattering energy $E$. The temperature is chosen as $T=0.2$ GK and the BN environment densities are varied from $\rho/\rho_0 = 0$ to $10^{-2}$.}
\label{Ffig}
\end{center}
\end{figure}

We consider now the effect of the environment BN on the neutron capture rates. In Fig.~4, we show the integrand, denoted by $f$ in the following, in Eq.~(\ref{rate2})  (a) for $^{58}$Ni and (b) for $^{78}$Ni,  as a function of the neutron scattering energy $E$. The temperature is chosen as $T=0.2$ GK and the BN environment densities are varied from $\rho/\rho_0 = 0$ to $10^{-2}$.
In each panel, the solid, long-dashed, dashed, dotted, dash-dotted, and dash-dot-dotted lines, respectively, correspond to
$\rho/\rho_0=0$, $5\times10^{-4}$, $10^{-3}$, $2\times10^{-3}$, $5\times10^{-3}$, and $10^{-2}$. The five finite densities yield for the energy shift due to the environment $\varepsilon_{\rm EMP}$ of about 6, 20, 60, 200, and 500~keV in that order.
At low finite densities, $\varepsilon_{\rm EMP}$ reduces $\sigma$ quite significantly in the low energy region, while the weighting factor $w\equiv \sqrt{E(E+\varepsilon_{\rm EMP})}$ increases. The change in $\sigma$ is saturated at a certain density because the energy dependence of $\sigma$ for $E \ga 200$ (100)~keV is quite small for the neutron capture by $^{58}$Ni ($^{78}$Ni) in our model. On the other hand, $w$ keeps increasing as the density becomes higher. Thus, the net effect of $\varepsilon_{\rm EMP}$ on $f$ is determined by the energy dependence of $\sigma$, particularly at low energies.

One sees from Fig.~4 that for the neutron capture by $^{58}$Ni, $\varepsilon_{\rm EMP}$ decreases $f$ for $0<\rho \la 5\times 10^{-3}\rho_0$ and increases $f$ for $\rho \ga 10^{-2}\rho_0$. On the other hand, for the neutron capture by $^{78}$Ni, $f$ changes very little for $\rho \la 2\times 10^{-3}$. This suggests that $\sigma$ for $^{78}$Ni follows well the $1/v$ law in the energy region shown in Fig.~4. It should be noted that we adopt a global optical potential that is not tested for the neutron-rich nucleus $^{78}$Ni. To draw a definite conclusion, more accurate and reliable evaluation of $\sigma$ at low energies will crucially be important.

\begin{figure}[htbp]
\begin{center}
\includegraphics[width=0.48\textwidth]{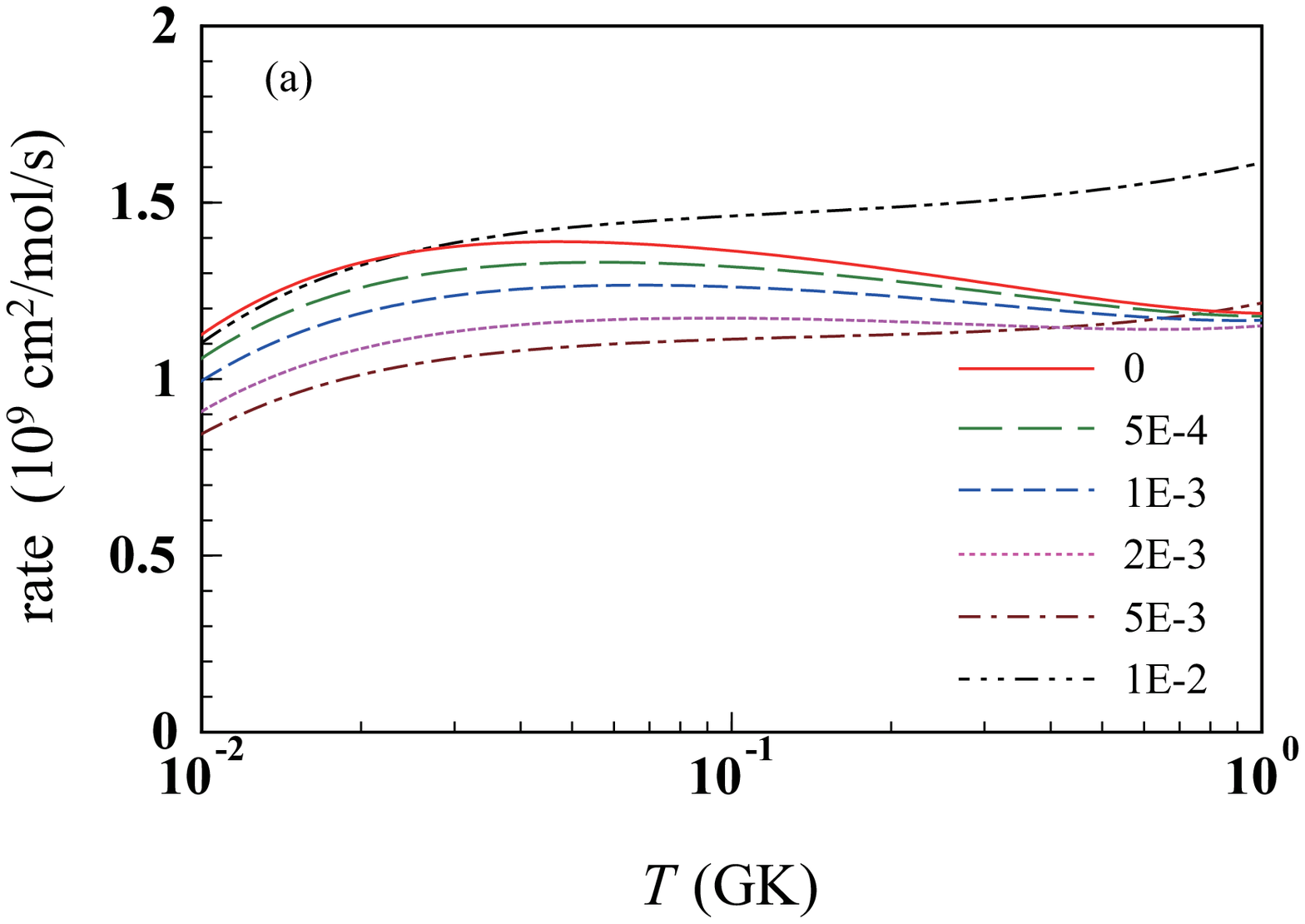}
\includegraphics[width=0.48\textwidth]{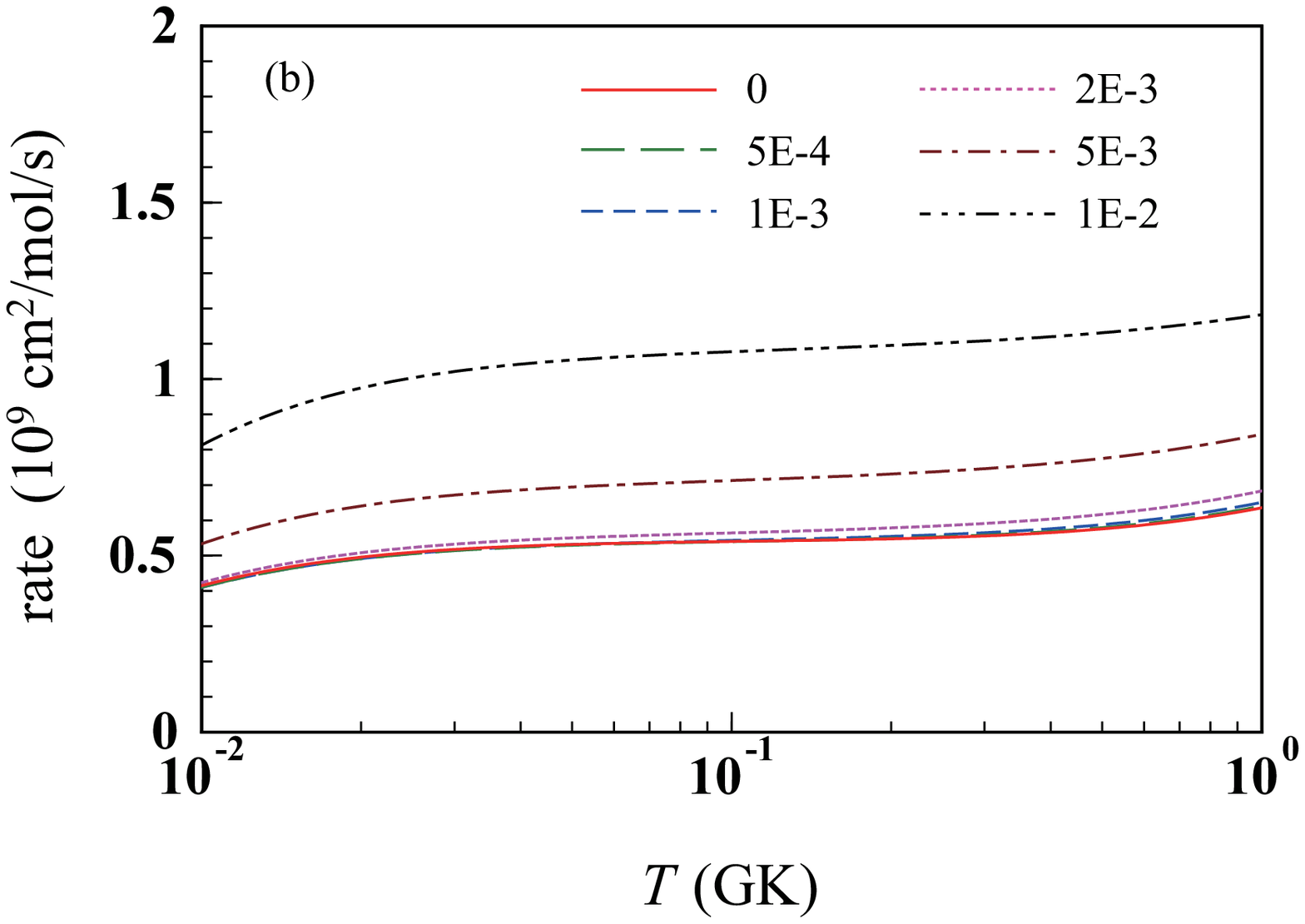}
\caption{Reaction rate $\left\langle \sigma v\right\rangle$ for $n$-${}^{58}$Ni and $n$-${}^{78}$Ni changes due to the energy shift $\varepsilon_{\rm EMP}$. These rates are displayed  in Figs.~5(a) and (b), respectively. The horizontal axis is the temperature in units of GK ($10^9$ K) and the correspondence between each line and the density is the same as in Fig.~4.}
\label{rfig}
\end{center}
\end{figure}

Finally, we show that the reaction rate $\left\langle \sigma v\right\rangle$ for $n$-${}^{58}$Ni and $n$-${}^{78}$Ni can substantially be modified because of the energy shift $\varepsilon_{\rm EMP}$. These rates are displayed  in Figs.~5(a) and 5(b), respectively. The horizontal axis is the temperature in units of GK ($10^9$ K) and the correspondence between each line and the density is the same as in Fig.~4. The behavior of the reaction rates can be understood by looking at the energy shift $\varepsilon_{\rm EMP}$ dependence of $f$ discussed above. The change in the rate at $\rho/\rho_0=10^{-2}$ is significantly large, which becomes even larger at higher density. It should be noted, however, that the central assumption made in this work, Eq.~(\ref{vemp}), will break down at some value of $\rho$ when it gets closer to the normal density $\rho_0$. A more careful treatment of the {\lq\lq}background'' neutrons that can be applied to $\rho \sim \rho_0$ will be interesting and important future work. We also remark here that the densities higher than $10^{-3}\rho_0$ will not directly be relevant to the standard r process. Nevertheless, if a nuclear reaction takes place at such densities and the ($n,\gamma$) $\leftrightarrow$ ($\gamma,n$) equilibrium is not realized, the neutron background effect will become crucial and must be included.

Another important observation is that the change in the reaction rate by the environment BN is expected for other nuclear reactions of various kinds. The energy shift $\varepsilon_{\rm EMP}$ will affect cross sections for proton-induced reactions in a very different way from that we have discussed here because the Coulomb penetrability dominates the cross section at low energies. In this case, the energy shift $\varepsilon_{\rm EMP}$ for $(S,T)=(1,0)$ must be evaluated. In addition, resonant capture processes can be affected by the energy shift $\varepsilon_{\rm EMP}$ significantly because it allows the reaction system to form resonant states at higher energies for which it is almost impossible to reach at the typical temperatures considered. In those investigations, however, one must be also aware of many uncertainties regarding nuclear properties such as strength function~\cite{Gor18}, statistical resonances~\cite{Roc17}, nuclear masses~\cite{Ma20}, and so on.

{\it Conclusions and final remarks.} We have considered the effect of environment neutrons in neutron capture reactions in dense environments. The effect is similar to that of electrons in dense stellar sites, such as those for which $\beta$-decay is blocked by the presence of the electrons in the final state, thus modifying the path of the r process. In contrast to electron environments, in a dense neutron environment the (n,$\gamma$) and ($\gamma$,n) are also influenced by Pauli blocking and our results indicate that this effect must be considered in reaction network calculations  for neutron star mergers involving equilibrium and non-equilibrium r processes. A fundamental difference from the treatment of electron-rich environment is that the neutron background will also have a decisive influence in thermal statistical equilibrium conditions.  Although in statistical equilibrium the (n,$\gamma$) and ($\gamma$,n) cross sections are not needed, the Pauli blocking of the emitted neutrons in the ($\gamma$,n) process will modify the Saha-like equations for statistical equilibrium. This might be easy to incorporate in detailed calculations by following a similar energy-shift method as proposed in this article.

\bigskip

{\it Acknowledgement.} We are grateful to Almudena Arcones and Moritz Reichert (TU-Darmstadt) for providing us with simulations for the density evolution of neutron star mergers and to Toshitaka Kajino (University of Tokyo and Beihang University) for beneficial discussions. This work has been supported by the  U.S. DOE grant DE-FG02-08ER41533 and Grants-in-Aid of the Japan Society for the Promotion of Science (Grant No. JP20K03971).

\end{document}